\documentclass[letterpaper,twocolumn,english,aps,pra,showpacs,preprintnumbers]{revtex4}
\usepackage[T1]{fontenc}
\usepackage[latin9]{inputenc}
\usepackage{graphicx}
\usepackage{amssymb}
\usepackage{babel}
\usepackage{babel}
\usepackage{babel}

\makeatletter
\@ifundefined{textcolor}{}
{
 \definecolor{BLACK}{gray}{0}
 \definecolor{WHITE}{gray}{1}
 \definecolor{RED}{rgb}{1,0,0}
 \definecolor{GREEN}{rgb}{0,1,0}
 \definecolor{BLUE}{rgb}{0,0,1}
 \definecolor{CYAN}{cmyk}{1,0,0,0}
 \definecolor{MAGENTA}{cmyk}{0,1,0,0}
 \definecolor{YELLOW}{cmyk}{0,0,1,0}
 }
\makeatletter
\@ifundefined{textcolor}{}
{
 \definecolor{BLACK}{gray}{0}
 \definecolor{WHITE}{gray}{1}
 \definecolor{RED}{rgb}{1,0,0}
 \definecolor{GREEN}{rgb}{0,1,0}
 \definecolor{BLUE}{rgb}{0,0,1}
 \definecolor{CYAN}{cmyk}{1,0,0,0}
 \definecolor{MAGENTA}{cmyk}{0,1,0,0}
 \definecolor{YELLOW}{cmyk}{0,0,1,0}
 }
\makeatother
\makeatother

\begin{document}

\title{Thermalization and temperature distribution in a driven ion chain}
\author{G.-D. Lin and L.-M. Duan}
\affiliation{Department of Physics and MCTP, University of Michigan, Ann Arbor, Michigan
48109}
\date{\today }

\begin{abstract}
We study thermalization and non-equilibrium dynamics in a dissipative
quantum many-body system --- a chain of ions with two points of the chain
driven by thermal bath under different temperature. Instead of a simple
linear temperature gradient as one expects from the classical heat diffusion
process, the temperature distribution in the ion chain shows surprisingly
rich patterns, which depend on the ion coupling rate to the bath, the
location of the driven ions, and the dissipation rates of the other ions in
the chain. Through simulation of the temperature evolution, we show that
these unusual temperature distribution patterns in the ion chain can be
quantitatively tested in experiments within a realistic time scale.
\end{abstract}

\maketitle

\section{Introduction}

Many-body non-equilibrium dynamics has raised significant interest in recent
years, in particular with connection to the atomic experiments where
far-from-equilibrium phenomena can be conveniently investigated due to the
long relaxation time of these systems \cite{1,2}. For instance, the dynamics
after a quantum quench has been studied in several model Hamiltonians \cite%
{3,4,5,6}. The physics becomes even richer if we add engineered dissipation
to the underlying system, which is possible to realize in atomic
experiments. Several interesting effects have been analyzed recently from
interplay between interaction and dissipation in cold atomic gas \cite{7,8}.

Motivated by this line of research, in this paper we study
thermalization and temperature distribution in a linear ion chain,
with two points of the chain driven respectively by a heating and
a cooling thermal bath. In a sense, the configuration here is an
analog of a classical example in thermodynamics --- the heat
propagation in a bar with its two ends fixed at different
temperature. In contrast to a simple linear temperature gradient
as one sees in a classical bar, we find that the temperature
distribution in a driven ion chain shows very rich behaviors in
its steady states: first, the temperature distribution depends
critically on the ratio between the driving speed and the
interaction rate in the ion chain. In the weak driving region, the
temperature distribution is non-monotonic across the ion chain and
shows a strange mirror effect. In the strong driving region,
instead of a linear temperature distribution, all the ions between
the two driven ones are stabilized to a constant temperature which
is in the middle of the two bath temperatures. Second, we find
that the bath which drives the ion with a medium coupling rate
plays a more effective role in determining the temperature of the
other ions in the system. When the bath drives the ion strongly,
it has little influence on temperature of the other ions in the
chain, which is pretty counter-intuitive. To see the unusual
phenomena predicted in this paper, we discuss the requirements for
an experimental observation and calculate the time dynamics to
achieve the steady-state temperature distribution. The qualitative
features of the steady-state temperature distribution patterns are
pretty insensitive to the size of the ion crystal and they show up
already in a pretty small ion chain with less than ten ions. The
realization time to the steady state is realistic for observation
compared with the current experimental time scale.

The paper is arranged as follows: In Sec. II we provide the theoretical
model and the calculation method. The main results of the temperature
distribution patterns under different circumstances are presented and
discussed in Sec. III. In Sec. IV, we discuss the time dynamics to achieve
the steady-state temperature distribution. Sec. V summarizes our major
findings.

\section{Model}

We consider a chain of ions along the axial $z$-direction coupled
with the Coulomb interaction and driven individually by a thermal
bath. The thermal bath can be provided, for instance, by cooling
or heating laser beams shined on each ion. Both the driving rate
and the effective bath temperature can be tuned by controlling the
intensity and the detuning of these laser beams. The oscillation
of the $i$th ion around its equilibrium position is described by
the coordinate and the momentum operators $x_{i},p_{i}$, with
the dynamics determined by the Heisenberg-Langevin equation:%
\begin{equation}
\left\{
\begin{array}{ccl}
\dot{x}_{i} & = & p_{i} \\
\dot{p}_{i} & = & -\sum_{j}A_{ij}x_{j}-\gamma _{i}p_{i}+\sqrt{2\gamma _{i}}%
\zeta _{i}(t)%
\end{array}%
\right. .  \label{eq:Langevin}
\end{equation}%
where $A_{ij}$ denotes the coupling matrix between the ions, $\gamma _{i}$
is the driving rate of the bath, and $\zeta _{i}(t)$ denotes the
corresponding random force from the thermal bath. We consider in this paper
the ions' motion along the transverse $x$ direction, so the coupling matrix $%
A_{ij}$ is given by $A_{ii}=\omega _{x}^{2}-\sum_{j(\neq
i)}1/\left\vert z_{j}-z_{i}\right\vert ^{3}$ and
$A_{ij}=1/\left\vert z_{j}-z_{i}\right\vert ^{3}$ for $i\neq j$,
where $\omega _{x}$ is the transverse trapping frequency and
$z_{i}$ denotes the ion equilibrium position along the axial
direction. We take the ion spacing $d_{0}$ as the length unit,
$e^{2}/d_{0}$ as the energy unit, and $\sqrt{e^{2}/(md_{0}^{3})}$
as the frequency unit so that every quantities in the matrix
$A_{ij}$ become dimensionless. Note that with control of an
anharmonic trapping potential along the axial direction, one can
make the ion spacing uniform in a scalable trap \cite{9}. For the
conventional harmonic axial trap, the ion spacing is not uniform.
In that case, the equilibrium positions $z_{i}$ are determined
numerically with the given trapping potential, and we take the
unit $d_{0}$ as the largest spacing in the ion chain. The
temperature distribution pattern that will be shown in Sec. III is
insensitive to the details of the axial trapping potential, and
for most of the calculations in the following, we assume the ion
spacing is uniform for simplicity (with exceptions in Sec. III for
consideration of the time dynamics in a harmonic trap). For
independent Markovian bath, the random force $\zeta _{i}(t)$ can
be expressed as $\zeta _{i}=-i\sqrt{\omega
_{i}/2}(b_{i}-b_{i}^{\dagger })$ with the bosonic field operator
$b_{i}(t)$
satisfying $\bigl\langle b_{i}^{\dagger }(t_{1})b_{j}(t_{2})\bigr\rangle%
=T_{i}^{B}\delta _{ij}\delta (t_{1}-t_{2})$, where $T_{i}^{B}$ is the
average phonon number which characterizes the temperature of the bath, and $%
\omega _{i}\triangleq \sqrt{A_{ii}}$ is the local oscillation frequency of
the $i$th ion by fixing all the other ions in their equilibrium positions.
For typical linear ion traps with strong transverse confinement, we have $%
\omega _{i}\approx \omega _{x}$. The correlation of $\zeta _{i}(t)$ is then
given by $\bigl\langle\zeta _{i}(t_{1})\zeta _{j}(t_{j})\bigr\rangle=\omega
_{i}(T_{i}^{B}+1/2)\delta _{ij}\delta (t_{1}-t_{2})$.

The Langevin equation (\ref{eq:Langevin}) can be solved exactly
through
diagonalization, with the solution formally expressed as%
\begin{equation}
\mathbf{q}(t)=e^{-\Omega t}\mathbf{q}(0)+\int_{0}^{t}d\tau e^{\Omega(\tau-t)}%
\mathbf{\eta}(\tau),
\end{equation}
where $\mathbf{q}\triangleq(x_{1},x_{2},...;p_{1},p_{2},...)^{\intercal}=%
\left[%
\begin{array}{c}
\left\{ x_{i}\right\} \\
\left\{ p_{i}\right\}%
\end{array}%
\right]$, $\eta(t)\triangleq\left[%
\begin{array}{c}
\left\{ 0\right\} \\
\left\{ \sqrt{2\gamma_{i}}\zeta_{i}\right\}%
\end{array}%
\right]$, and $\Omega\triangleq\left[{\normalcolor
\begin{array}{cc}
{\normalcolor 0} & {\normalcolor -I} \\
{\normalcolor \bigl[A_{ij}\bigr]} & \bigl[\gamma_{i}\delta_{ij}\bigr]%
\end{array}%
}\right]$ is a $2N\times2N$ matrix, which can be diagonalized as $\bigl[%
U^{-1}\Omega
U\bigr]_{\alpha\beta}=\lambda_{\alpha}\delta_{\alpha\beta}$. From
this formal solution, we obtain the variance of the operators
$x_{i}$ and $p_{i}$ as
\begin{eqnarray}
\bigl\langle q_{\mu}^{2}\bigr\rangle & = &
\sum_{i=1}^{N}\sum_{\alpha,\beta=1}^{2N}U_{\mu\alpha}U_{\mu\beta}\Biggl(%
e^{-(\lambda_{\alpha}+\lambda_{\beta})t}\biggl[\bigl\langle x_{i}^{2}(0)%
\bigr\rangle U_{\beta i}^{-1}U_{\alpha i}^{-1}  \nonumber \\
& + & \bigl\langle p_{i}^{2}(0)\bigr\rangle U_{\beta,i+N}^{-1}U_{%
\alpha,i+N}^{-1}\biggr]  \label{eq:Variance} \\
& + & \Bigl(1-e^{-(\lambda_{\alpha}+\lambda_{\beta})t}\Bigr)\frac{%
2\omega_{i}\gamma_{i}(T_{i}^{B}+\frac{1}{2})}{\lambda_{\alpha}+\lambda_{%
\beta}}U_{\beta,i+N}^{-1}U_{\alpha,i+N}^{-1}\Biggr),  \nonumber
\end{eqnarray}
where $\mu=1,2,...,N$ correspond to the $x$-operators and $%
\mu=N+1,N+2,...,2N $ correspond to the $p$-operators. The temperature of
each ion is similarly characterized by the average phonon number of its
thermal oscillation, so the temperature of the $i$th ion as a function of
time $t$ is represented by $T_{i}(t)=\frac{1}{2}\bigl(\omega_{i}\bigl\langle %
x_{i}^{2}(t)\bigr\rangle+\bigl\langle p_{i}^{2}(t)\bigr\rangle/\omega_{i}-1%
\bigr)$. As long as $\gamma_{i}$ is nonzero for some ions, each eigenvalue $%
\lambda_{\alpha}$ has a positive real part due to coupling of all
the ions, and the system approaches to a steady state as
$t\longrightarrow\infty$. The steady state temperature for each
ion is denoted by $T_{i}^{s}\triangleq
T_{i}(t\longrightarrow\infty)$.

\section{Steady-State Temperature Distribution}

\begin{figure}[tbp]
\includegraphics[width=7cm]{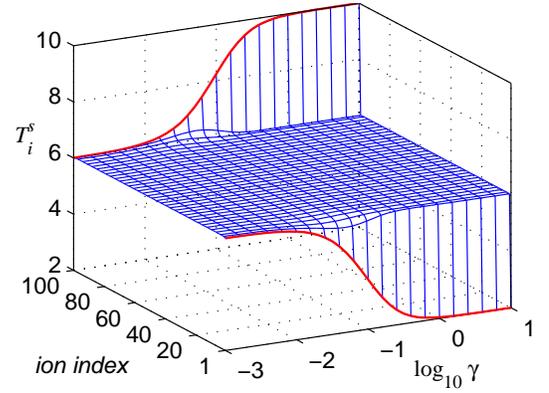}
\caption{(Color online) The steady-state ($t\rightarrow\infty$)
temperature distribution $T_{i}^{s}$ (measured by the mean thermal
phonon number) of a uniform ion chain with $N=100$, where the
$1^{\mbox{st}}$ and the last ($100^{\mbox{th}}$) ions are driven
respectively by a cooling (at temperate $T_{1}^B=2$) and a heating
(at temperate $T_{2}^B=10$) thermal bath under varying driving
rate $\protect\gamma$. The temperature of the driven ions
($1^{\mbox{st}}$ and $100^{\mbox{th}}$) are indicated by the thick
curves. The other ions are assumed to be isolated from the bath,
and the transverse trapping frequency $\protect\omega_{x}=$10 in
the unit of $\sqrt{e^{2}/(md_{0}^{3})}$, where $d_0$ is the ion
spacing. } \label{fig:gamma_transition}
\end{figure}

In analogy to the example of heat propagation in a conducting bar, we
consider a long ion chain with the two edge ions driven by different thermal
bath with temperature $T_{1}^{B}$ and $T_{2}^{B}$ (with $T_{1}^{B}<T_{2}^{B}$%
), respectively. For simplicity, we assume the corresponding driving rate $%
\gamma _{1}=\gamma _{2}=\gamma $ and all the ions in the middle have no
dissipation. After all the ions attain the steady state, the temperature
distribution $T_{i}^{s}$ across the chain is shown in Fig. \ref%
{fig:gamma_transition} for $N=100$ ions under different driving rates $%
\gamma $. Apparently, the temperature distribution does not follow a linear
gradient. There are three regions for the distribution of $T_{i}^{s}$,
depending on the ratio between the driving rate $\gamma $ and the ion
interaction rate. Note that in our unit the interaction energy between the
neighboring ions is of the order of unity. If $\gamma \gg 1$, the
dissipation is much faster than the energy propagation in the chain, and
without surprise the two edge ions have temperature basically fixed by their
corresponding bath temperature $T_{1}^{B}$ and $T_{2}^{B}$. However, it is
surprising that all the other ions in the middle approach almost the same
temperature given by $T_{i}^{s}\simeq (T_{1}^{B}+T_{2}^{B})/2$ for $%
i=2,3,\cdots ,99$ in this case. The temperature does not fall down
gradually from the hot end to the cold end as in the classical
heat propagation problem, but takes a sharp jump right from the
driven ion to the next one and then keeps constant over the whole
chain. In the opposite limit of weak driving with $\gamma \ll 1$,
the phonon propagation is faster than the bath driving, and all
the ions approach the same temperature. It sounds that the
temperature distribution in this limit resembles a classical
thermal
equilibrium, however, this picture is not true. The example in Fig. \ref%
{fig:gamma_transition} represents an exception instead of a rule
where we put the cooling and heating ions exactly at the symmetric
positions of an ion chain. As we will see in the following, when
we shift the position of one of the ions to break the reflection
symmetry, the temperature distribution in the weak driving limit
has strange features. It is even not monotonic across the ion
chain, in sharp contrast with the distribution from the diffusion
process. Between these two limiting regions there lies a
transition region with $0.01<\gamma <1$, where the temperature of
the edge ions gradually approach the corresponding bath
temperature. Note that the temperature for the ions next to the
driven ones follow a non-monotonic curve as one changes $\gamma $,
although such variation is pretty small.

\begin{figure}[tbp]
\includegraphics[width=7cm]{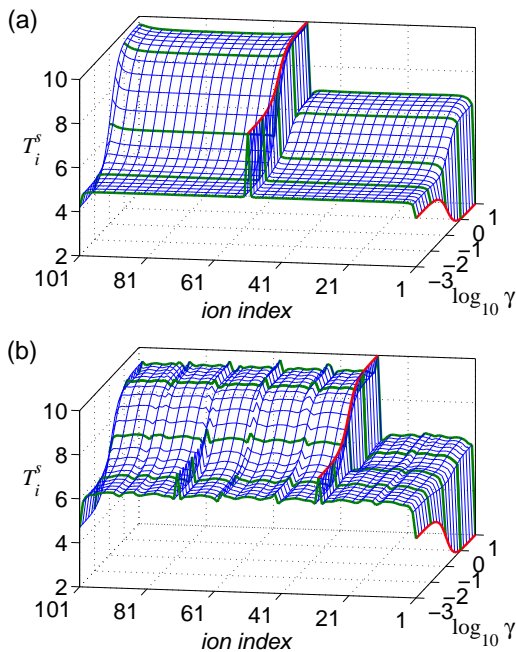}
\caption{(Color online) The steady-state temperature distribution
for $N=101$ ions, where the cooling bath (at $T_{1}^{B}=2$)
remains at the $1^{\mbox{st}}$ ion and the heating bath (at
$T_{2}^{B}=10$) moves to (a) the middle ($51^{\mbox{st}}$) ion,
and (b) the $30^{\mbox{th}}$ ion. The temperature of the driven
ions are indicated by thick (red) curves along the axis of the
driving rate $\protect\gamma$. The other parameters are the same
as in Fig. 1. The temperature profiles are highlighted by
different thick (green) curves along the \textsf{\textit{ion
index}} axis to represent $\protect\gamma=10^{-3},10^{-2},0.1,1,10$.
For small $\protect%
\gamma$, note that the temperature profiles both in (a) and (b)
are found to have the mirror effect, i.e., a reflection symmetry
of the distribution with respect to the middle ($51^{ \mbox{st}}$)
ion.} \label{fig:mirror}
\end{figure}

To break the position symmetry of the cooling and the heating ions, we fix
the cooling ion still at the edge, but move the heating ion inside the
chain. Fig. \ref{fig:mirror}(a) shows an example of the temperature
distribution where the heating ion is right at the center of the chain. When
$\gamma\gg1$, the heating ion separates two plateaus in the temperature
distribution: a lower temperature region between the heating and the cooling
ions and a higher temperature region on the other side of the heating ion.
As $\gamma$ decreases, the two plateaus smoothly descend to the same
temperature while the heating ion remains as a peak in the temperature
distribution. When $\gamma\ll1$, the temperature distribution has a
reflection symmetry with respect to the chain center. A dip in temperature
appears on the free edge of the chain (the left side of Fig. \ref{fig:mirror}%
(a)), mirroring the cooling ion on the other side of the chain (the mirror
effect). The temperature distribution is apparently non-monotonic in this
case. The mirror effect exists for other positions of the heating ion. For
instance, Fig. \ref{fig:mirror}(b) shows the temperature distribution where
the distance between the heating and the cooling ions is about one third of
the chain length. Both the cooling and the heating ions have their mirror
images in the temperature distribution in the weak driving limit.

\begin{figure}[bp]
\includegraphics[width=7cm]{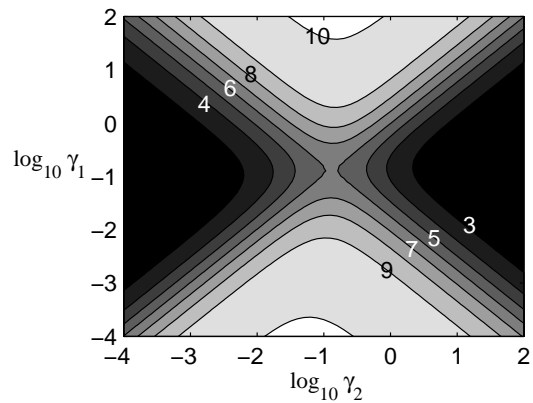}
\caption{The middle-ion steady-state temperature $T_{m}^s$ for the
configuration shown in Fig. 1, as a function of two independent
bath driving rates $\protect\gamma_{1}$ and $\protect\gamma_{2}$.
The dark (light) regions correspond to low (high) temperatures,
and the numbers there indicate different temperature contours. The
parameters are the same as in Fig. \protect\ref
{fig:gamma_transition} expect
that $\protect\gamma_{1}$ associated with the cooling bath(at $%
T_{1}^{B}=2$) and $\protect\gamma_{2}$ associated with the heating
bath (at $T_{2}^{B}=10$) are varied independently. }
\label{fig:2gamma_map}
\end{figure}

In the above calculation, we assumed $\gamma _{1}=\gamma _{2}$ for
the cooling and the heating ions. When these two driven ions are
put at the edge, the middle ions are at a constant temperature
$T_{m}$ which is exactly the average of the two bath temperatures.
If the driving rate $\gamma _{1}\neq \gamma _{2}$, we may expect
that the bath which drives the ion more strongly plays a larger
role in determining the temperate $T_{m}$ of the middle segment
ions. This expectation, however, turns out to be not true. To show
that, we map out the middle segment temperature $T_{m}$ in Fig. \ref%
{fig:2gamma_map} as a function of $\gamma _{1}$ and $\gamma _{2}$. The
figure shows that $\gamma _{i}\approx 0.1$ is the optimal driving rate for
which the corresponding bath has the largest influence on the temperature $%
T_{m}$. At this optimal value, the driving rate is comparable with the ion
interaction rate in term of the order of magnitude. The middle segment
temperature is close to the temperature of the bath that drives the ion at
the optimal rate. When the driving gets too strong or too weak, the bath
plays little role in determining the temperature of the other ions in the
chain. This result has important implication for the sympathetic cooling
\cite{10,11,12}: instead of fast cooling of the ancilla ions, cooling at a
moderate optimal rate is more efficient in reducing the temperature of the
computational ions.

\begin{figure}[tbp]
\includegraphics[width=7cm]{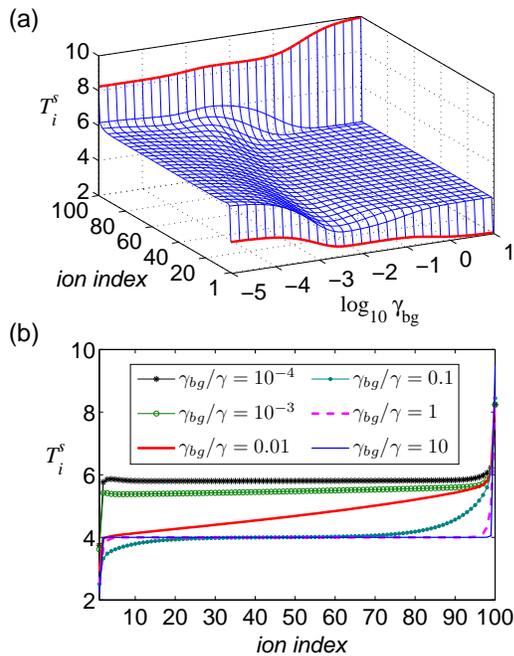}
\caption{(Color online) (a) The steady-state temperature
distribution across the ion chain, when the other ions in the
chain (except the two driven ones at the edge) are coupled to a
background thermal bath at temperature $T_{bg}^{B}=4$ with a
coupling rate $\protect\gamma_{2}=...=\gamma_{99}=\protect%
\gamma_{bg}$. (b) The temperature profiles at different
$\protect\gamma_{\mbox{bg}}$ (cross-section view of (a)). We find
that the temperature profile for the main part of the chain,
excluding the two edge ions, approaches a linear distribution
at an intermediately small $\protect\gamma_{\mbox{bg}}\simeq10^{-2}%
\protect\gamma$.} \label{fig:background_decoh}
\end{figure}

So far we have neglected dissipation of the middle segment ions. Now, apart
from the cooling and the heating bath (with temperature $T_{1}^{B}$ and $%
T_{2}^{B}$, respectively) attached to the two edge ions, we assume all the
middle ions are coupled to a background bath with temperature $T_{bg}^{B}$ ($%
T_{1}^{B}<T_{bg}^{B}<T_{2}^{B}$) at a coupling rate $\gamma_{bg}$. The final
temperature distribution of the ions is shown in Fig. \ref%
{fig:background_decoh} under different background coupling rates $%
\gamma_{bg} $. In this calculation, we fix $\gamma_{1}=\gamma_{2}=\gamma=0.1$%
. For tiny $\gamma_{bg}$, the temperature distribution shows no noticeable
difference compared with the case of $\gamma_{bg}=0$. However, as $%
\gamma_{bg}$ increases to a moderate value with $\gamma_{bg}/\gamma\sim0.01$%
, the temperature distribution of the middle segment ions has a clear linear
spatial gradient, resembling the linear temperature distribution of a
classical bar in the heat diffusion problem. As $\gamma_{bg}$ further
increases and gets close to $\gamma$, the temperature of the middle ions are
pined to the background bath temperature $T_{bg}^{B}$ as one expects and the
temperature gradient disappears again in this limit.

\section{Temperature evolution}

\begin{figure}[bp]
\includegraphics[width=7cm]{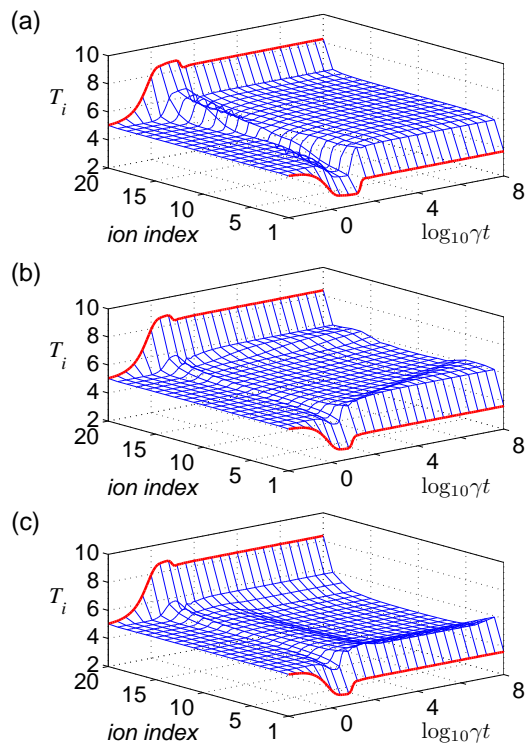}
\caption{(Color online) The  temperature relaxation dynamics for
an ion chain with $N=20$ in (a) a uniform anharmonic trap, (b) a
harmonic non-uniform trap, and (c) a harmonic trap with a
background coupling rate
$\protect\gamma_{bg}/\protect\gamma=10^{-3}$ and a
background temperature $T_{bg}^{B}=4$. The $1^{\mbox{st}}$ ion and
the $20^{\mbox{th}}$ ions are driven respectively by a cooling (at $%
T_{1}^{B}=2$) and a heating (at $T_{2}^{B}=10$) bath, with the
same driving rate $\protect\gamma_1=\gamma_2=\gamma=0.1$, as
indicated by the thick (red) curves. The initial temperature is
assumed to be  $T_{i}(t=0)=5$ for all the ions. The transverse
frequency $\protect\omega_{x}=10$ in the unit of
$\sqrt{e^{2}/(md_{0}^{3})}$, where $d_0$ is the largest spacing in
the ion chain (between the $1^{\mbox{st}}$ and the
$2^{\mbox{nd}}$, or the $19^{\mbox{th}}$ and the $20^{ \mbox{th}}$
ions) for the harmonic cases. In terms of real numbers, with a
typical choice of $d_{0}=10\protect\mu\mbox{m}$ for 20 ytterbium
ions in a harmonic trap, the trapping frequencies for the
transverse and the axial traps are given respectively by
$\omega_x \approx 2\protect\pi\times1.4\mbox{MHz}$ and $\omega_z \approx 2\protect\pi\times76.5\mbox{kHz}$%
, and the driving rate $\gamma\approx 14 \mbox{kHz}$.}
\label{fig:dynamics}
\end{figure}

We note that the temperature distributions shown in this paper are
insensitive to the size of the ion system. We have checked the
temperature distribution of the ions with the size of the chain
varying from ten to a few hundreds of ions and noticed no qualitative
change of the distribution pattern. To experimentally test these
unusual temperature distribution patterns, one can start with a
small system of a few ions that are within the reach of the
current experimental technology. The final temperature (the mean
phonon number of the ion motion) can be detected, for instance, by
measuring the scattering sidebands of a laser beam through a ion.
The asymmetry in the blue and the red sidebands and their ratio
give direct inference of the thermal phonon number of the ion
motion \cite{13,14,15}.

To probe the properties of the steady states, we need to know the
time scale to approach these steady states. Fig.
\ref{fig:dynamics} shows the relaxation dynamics for $20$ ions in
either a harmonic or an anharmonic uniform trap \cite{9}. For a
uniform trap (in Fig. \ref{fig:dynamics}a), the calculation shows
that there are two time scales in the thermalization. First, with
a time scale $t_{1}\sim 1/\gamma $, the temperature of the two
edge ions quickly approaches the corresponding bath temperature.
During this step, the temperature of the middle segment ions only
changes slightly, and the change gets smaller as one moves away
from the driven ions. After that, a longer time scale $t_{2}$ sets
in, representing the interaction driven thermalization process.
All the ions gradually approach the steady-state temperature. Note
that the temperature of the edge ions (as a well as the other ions
that are close to the two driven ones) does not follow a monotonic
evolution curve. Instead, it first comes pretty close to the
corresponding bath temperature and then is dragged back toward its
steady state value. The value of the second time scale $t_{2}$
increases with the system size (roughly linearly) and $t_{2}\sim
40/\gamma $ for $20$ ions. For a ytterbium ion
($^{171}\mbox{Yb}^{+}$) chain with spacing 10 $\mu $m, $t_{2}$ is
about $3$ ms, which is a pretty reasonable time scale for
experiments. For a harmonic trap (in Fig. \ref{fig:dynamics}b),
the basic feature is similar except that the second time scale
$t_{2}$ gets site dependent. The edge ions and their neighbors
approach the corresponding steady-state temperature with a time
scale that is comparable with the case of a uniform chain.
However, as one moves away from the edge, the thermalization time
gets much longer with an exponential increase. For the middle ion,
the thermalization is not finished yet with $t\sim 10^{10}/\gamma
$. With such an extremely long time scale, of course one can not
neglect the small dissipation of the other ions to the background
bath. If we take into account a small background dissipation,
e.g., with a rate $\gamma _{bg}/\gamma \sim 10^{-3}$ as shown in
Fig. \ref{fig:dynamics}c, the long tale in the temperature
evolution is completely gone for the harmonic trap. Now, the
temperature of all the ions approach their steady values within a
time scale that is comparable with the uniform case, although the
steady-state temperature of the middle segment ions is dragged
down a bit toward the background bath temperature.

\section{Conclusion}

In summary, we have shown that the steady state temperature
distribution of a driven ion chain shows surprisingly rich
patterns. Many features of these patterns are unexpected, and they
are in sharp contrast with the simple linear temperature gradient
as one sees in the classical heat diffusion problem. Our
calculation is based on exact numerical methods with no unreliable
approximations, so we believe all the unusual temperature
distribution patterns revealed by this calculation will show up in
experiments, although we still lack intuitive explanation of many
of these features. We also investigate the relaxation dynamics and
the time scale to reach the steady state, and show that these
patterns should be observable under a realistic time scale in a
small system that is within the current experimental reach.

\begin{acknowledgments}
We thank C. Monroe and Y.-J. Han for helpful discussions. This
work is supported by the DARPA OLE Program under ARO Award
W911NF-07-1-0576, the ARO MURI program, the IARPA, and the AFOSR
MURI program.
\end{acknowledgments}

\end{document}